\documentclass{emulateapj}
\usepackage{apjfonts}
\journalinfo{The Astrophysical Journal, 679, 2008 June 1, in press}

\shorttitle{HYPERDIFFUSION AS A CORONAL HEATING MECHANISM}
\shortauthors{VAN BALLEGOOIJEN AND CRANMER}

\begin{document}

\title{Hyperdiffusion as a Mechanism for Solar Coronal Heating}
\author{A. A. van Ballegooijen and S. R. Cranmer}
\affil{Harvard-Smithsonian Center for Astrophysics,
60 Garden Street, Cambridge, MA 02138, USA}
\email{vanballe@cfa.harvard.edu}

\begin{abstract}
A theory for the heating of coronal magnetic flux ropes is developed.
The dissipated magnetic energy has two distinct contributions:
(1) energy injected into the corona as a result of granule-scale,
random footpoint motions, and (2) energy from the large-scale,
nonpotential magnetic field of the flux rope. The second type of
dissipation can be described in term of hyperdiffusion, a type of
magnetic diffusion in which the helicity of the mean magnetic field is
conserved. The associated heating rate depends on the gradient of the
torsion parameter of the mean magnetic field.  A simple model of an
active region containing a coronal flux rope is constructed.  We find
that the temperature and density on the axis of the flux rope are
lower than in the local surroundings, consistent with observations
of coronal cavities. The model requires that the magnetic field in
the flux rope is stochastic in nature, with a perpendicular length
scale of the magnetic fluctuations of order 1000 km.
\end{abstract}


\newcommand{\bnabla}{\mbox{\boldmath $\nabla$}}
\newcommand{\taup}{\tau_{\rm ph}}
\newcommand{\uphot}{u_{\rm ph}}
\newcommand{\bphot}{B_{\rm ph}}
\newcommand{\lenp}{\lambda_{\rm turb}}
\newcommand{\bphi}{\mbox{\boldmath $\hat{\phi}$}}

\section{Introduction}

In this paper we consider the heating of the solar corona in regions
where the large-scale magnetic field deviates significantly from a
(current-free) potential field. Evidence for such nonpotential
structures comes from a variety of sources. Measurements of
photospheric vector fields in active regions often show deviations
from the potential field \citep[e.g.,][] {Gary1987, Pevtsov1995,
Pevtsov1997, Leka1996}. X-ray observations show {\it sigmoids}, which
are bright S-shaped or inverse S-shaped structures inside or between
active regions \citep[e.g.,][] {Acton1992, Rust1996, Canfield1999,
Sterling2000, Gibson2002}. These sigmoids contain coronal loops that
are highly sheared with respect to the photospheric polarity inversion
line (PIL), and are are often associated with H$\alpha$ filaments
\citep[e.g.,][] {Manoharan1996, Pevtsov2002, Gibson2006a}.
\citet{Schmieder1996} proposed a differential magnetic
shear model in which the degree of shear decreases with height in the
observed active region. \citet{Schrijver2005} studied the
nonpotentiality of active regions using extreme-ultraviolet (EUV)
images from the {\it Transition Region and Coronal Explorer} (TRACE).
They found that significant nonpotentiality occurs when new magnetic
flux has recently emerged into the corona, or when rapidly evolving,
opposite polarity flux concentrations are in close contact.
\citet{Mandrini2000} tested various models of coronal heating using
potential, linear force-free, and magnetostatic models. Of course,
nonpotential magnetic fields also play an important role in solar
flares and coronal mass ejections \citep[see] [and references therein]
{Fan2007}.

The nonpotential structures in the corona have been modeled as {\it
flux ropes} in which the field lines twist about an axial field line
\citep[e.g.,][]{Rust1994, Low1995, Aulanier1998, Titov1999, Amari2003,
Magara2001, Kliem2004, Kusano2005, Aulanier2005, Gibson+etal2006a}.
The presence of a coronal flux rope implies the existence of electric
currents that pass through the photosphere and into the corona.  The
corona of an active region is a magnetically dominated plasma.
Therefore, except during a flare or eruption, the corona is nearly
force free, $\bnabla \times {\bf B} \approx \alpha {\bf B}$, where
${\bf B} ({\bf r})$ is the magnetic field and $\alpha ({\bf r})$ is
the so-called torsion parameter. A magnetic field with this property
is called a nonlinear force-free field (NLFFF). Many authors have
developed NLFFF models of active regions, either by extrapolating
photospheric vector fields into the corona \citep[e.g.,][]
{Wheatland2000, Bleybel2002, Regnier2002, Wiegelmann2004, Valori2005,
Wheatland2006, Regnier2007}, or using a flux rope insertion method
\citep[][] {vB2004, Bobra2008}.

Active regions exhibit transient brightenings at EUV and soft X-ray
wavelengths on time scales of a few minutes \citep[e.g.,][]
{Shimizu1992, Shimizu1995, Porter1995, Berghmans1999, Berghmans2001}.
These single- or multi-loop
brightenings are thought to be due to the release of magnetic energy
by small reconnection events in the complex coronal field of an active
region. This may be a two-step process: magnetic energy may first be
converted into kinetic energy of reconnection outflows, waves or
energetic particles, and then be converted into thermal energy of the
plasma \citep{Longcope2004}. It seems likely that X-ray sigmoids are
also heated by such reconnection events, with many events occuring
simultaneously at different sites within the sigmoid.  The events may
occur so frequently that the coronal plasma does not have time to cool
significantly between heating events, keeping the plasma at
temperatures of 3-5 MK \citep[e.g.,][]{Klimchuk1995, Cargill2004}.
This would explain why sigmoids generally do not show up in EUV
spectral lines formed at 1-1.5 MK.

What is the origin of the energy released in small reconnection
events? One possibility is that the energy is injected into the
corona by small-scale random footpoint motions, which lead to twisting
and braiding of coronal field lines \citep[e.g.,][] {Parker1972,
Parker1988}. However, in the highly sheared magnetic field of a
sigmoid there is another possibility, namely, that the energy for
coronal heating is derived from the energy of the flux rope itself.
In this case the energy is already stored in the corona and does not
need to be injected at the photospheric footpoints. Generally, a large
amount of free energy is available in the coronal flux rope ($\sim
10^{32}$ erg), sufficient to explain both the heating of the coronal
plasma and the occurrence of flares and coronal mass ejections.

\begin{figure}
\epsscale{1.1}
\plotone{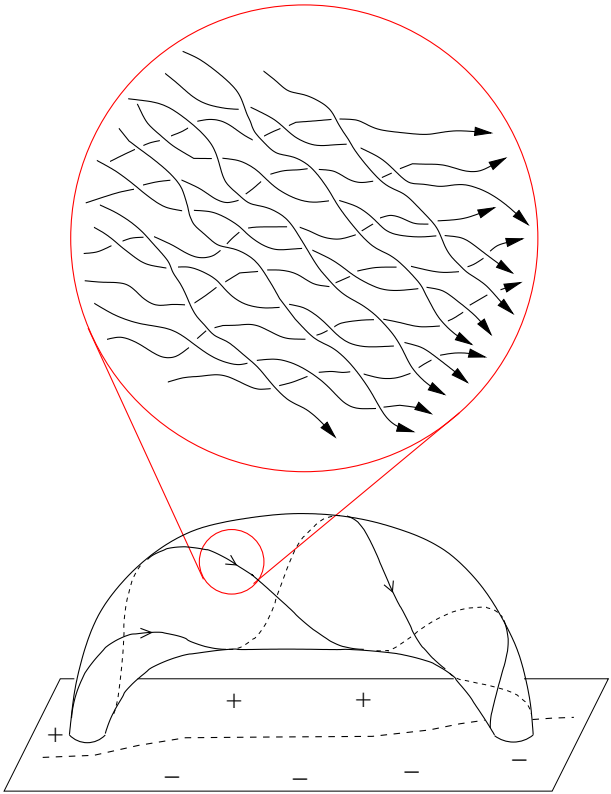}
\caption{Stochastic magnetic fields in a coronal flux rope.}
\label{fig1}
\end{figure}

The second scenario leads us to propose a new theory for the heating
of coronal flux ropes. The basic ideas are illustrated in Figure
\ref{fig1}. The lower part of the figure shows a flux rope in the
low corona. Inside the flux rope is a magnetic field that
runs more or less parallel to the flux rope axis. The flux rope is
held down by an overlying coronal arcade (not shown), which has a
magnetic field that runs perpendicular to the flux rope axis. Hence,
going from the inside to the outside of the flux rope the direction of
the mean magnetic field rotates through an angle of about 90
degrees. The upper part of the figure shows the fine-scale structure
of the magnetic field at the boundary between the flux rope and its
surroundings. The magnetic field has both mean and fluctuating
components, as indicated by the wiggles in the field lines. The total
magnetic field ${\bf B} ({\bf r})$ is assumed to be stochastic in
nature \citep[][]{Lazarian1999}. The magnetic fluctuations are assumed
to be produced by some unspecified MHD process that ultimately derives
its energy from the free magnetic energy of the coronal flux rope.
\citet{Strauss1988} proposed that turbulent fluctuations could be
produced by tearing modes in coronal current sheets, but the present
paper does not depend on that particular mechanism. We only assume
that reconnection occurs intermittently at many sites in the
stochastic field. The reconnection events produce outflows and
energetic particles that ultimately result in heating of the coronal
plasma.

The above process is an example of a ``dynamo'' process because it
involves turbulent diffusion of the mean magnetic field of the coronal
flux rope. Dynamo activity, defined as the amplification and/or
maintenance of large-scale magnetic fields, occurs in many
astrophysical and laboratory plasmas, and magnetic helicity evolution
plays an important role in such activity \citep[e.g.,][]{Blackman2006,
Subramanian2006}. \citet{Boozer1986} and \citet{Bhattacharjee1986}
considered turbulence in low-$\beta$ plasmas, and proposed that the
evolution of the mean magnetic field can be described in term of
{\it hyperdiffusion}, a type of magnetic diffusion in which the
magnetic helicity of the mean field is conserved.
\citet{Strauss1988} proposed that tearing modes can produce turbulence
in three-dimensional sheared magnetic fields, and showed that such
turbulence causes hyperdiffusion, which may explain both fast 
reconnection in solar flares and the heating of the solar corona.
\citet{Bhattacharjee1995} considered the constraints on the dynamo
mechanism from helicity conservation and the back-reaction of the
magnetic field on plasma dynamics. They showed that the results of
kinematic dynamo theory are strongly modified by the production of
hyperdiffusion in the mean field equation. Hyperdiffusion has also
been included in models for decaying active regions \citep[][]
{vB+Mackay2007}.

In the present paper we consider the plasma heating associated with
hyperdiffusion. At present, it is unclear to what extent stochastic
effects dominate the solar magnetic field. Our purpose here is not to
develop a complete theory of such fields, but rather to determine some
of the observable consequences of their existence. Therefore, more
detailed questions, such as how the stochastic field is produced and
the nature of the MHD instabilities, will remain unanswered. The main
goal is to demonstrate that the existence of stochastic fields is
consistent with existing observations of coronal heating in active
regions.

\section{Theory of Hyperdiffusion}

The theory of hyperdiffusion was first proposed by \citet{Boozer1986}
and \citet{Bhattacharjee1986}, and further developed by
\citet{Strauss1988}, \citet{Bhattacharjee1995}, \citet{Vishniac2001}
and others. In this section we summarize the results of these analyses
as they apply to the solar corona, i.e., when the plasma is
magnetically dominated. The total magnetic field ${\bf B} ({\bf r},t)$
is a sum of mean and fluctuating components, and is described in terms
of the vector potential ${\bf A} ({\bf r},t)$, where ${\bf B} \equiv
\bnabla \times {\bf A}$. The evolution equation for the vector
potential is
\begin{equation}
\frac{\partial {\bf A}} {\partial t} = - c {\bf E}  - \bnabla \Phi ,
\label{eq:E}
\end{equation}
where ${\bf E} ({\bf r},t)$ is the electric field, $\Phi ({\bf r},t)$
is the scalar potential, and $c$ is the speed of light. Inserting
Ohm's law and Ampere's law, we obtain
\begin{equation}
\frac{\partial {\bf A}} {\partial t} = {\bf v} \times {\bf B} -
\eta \bnabla \times {\bf B} - \bnabla \Phi , \label{eq:dadt}
\end{equation}
where ${\bf v} ({\bf r},t)$ is the plasma velocity, and $\eta$ is the
Ohmic diffusivity. We assume that $\eta \ll L^2/\tau$, where $L$ is
the system size and $\tau$ is the time scale for the evolution of the
mean magnetic field. Then reconnection occurs only in highly localized
regions (e.g., current sheets) that make up only a small fraction of
the volume. Following \citet{Vishniac2001}, we use the Coulomb gauge,
$\bnabla \cdot {\bf A} = 0$, so that the scalar potential is given by
\begin{equation}
\nabla^2 \Phi = \bnabla \cdot \left( {\bf v} \times {\bf B} - \eta
\bnabla \times {\bf B} \right) . \label{eq:Phi2}
\end{equation}
The magnetic helicity equation follows from equation (\ref{eq:dadt}):
\begin{equation}
\frac{\partial} {\partial t} \left( {\bf A} \cdot {\bf B} \right)
+ \bnabla \cdot {\cal H} = - 2 \eta ~ {\bf B} \cdot \bnabla \times
{\bf B}, \label{eq:heli}
\end{equation}
where $({\bf A} \cdot {\bf B})$ is the helicity density, and ${\cal H}
({\bf r},t)$ is the helicity flux:
\begin{eqnarray}
{\cal H} & \equiv & {\bf A} \times \left( {\bf v} \times {\bf B} -
\eta \bnabla \times {\bf B} + \bnabla \Phi \right) \nonumber \\
 & = & {\bf A} \times \left( -c {\bf E} + \bnabla \Phi \right) . 
\label{eq:H}
\end{eqnarray}
\citet{Berger1984} has shown that in the limit $\eta \ll L^2/\tau$
the right-hand side of equation (\ref{eq:heli}) can be neglected, so
this equation describes the conservation of magnetic helicity.
The relative magnetic helicity is defined by $H_R \equiv
\int [ {\bf A} \cdot {\bf B} - {\bf A}_p \cdot {\bf B}_p ] ~ dV$,
where subscript $p$ refers to the potential field, and the
integration includes the coronal and subsurface regions \citep[see][]
{Berger+Field1984}. Therefore, in the limit $\eta \ll L^2/\tau$,
if no helicity is injected into the corona at the photosphere or
ejected into the heliosphere, $H_R$ is constant in time.

Let ${\bf B}_0 ({\bf r},t)$ be the mean magnetic field smoothed over
some intermediate spatial scale $\ell$, and let ${\bf B}_1 ({\bf r},
t)$ be the fluctuating field on scales less than $\ell$. Similar
decompositions are made for other physical quantities. Applying the
smoothing operator to equations (\ref{eq:E}) and (\ref{eq:dadt}), we
obtain:
\begin{equation}
\frac{\partial {\bf A}_0} {\partial t} = 
- c {\bf E}_0 - \bnabla \Phi_0 \approx
{\bf v}_0^\prime \times {\bf B}_0 + {\cal E}_0 - \bnabla \Phi_0 ,
\label{eq:dadt0}
\end{equation}
where ${\bf A}_0$, $\Phi_0$, ${\bf E}_0$ and ${\bf v}_0^\prime$ are
the mean vector potential, scalar potential, electric field and plasma
velocity, and ${\cal E}_0 \equiv < {\bf v}_1 \times {\bf B}_1 >$ is
the ``mean electromotive force'' describing the effect of the
fluctuations on the mean magnetic field \citep[][]{Steenbeck1966}.
Here we neglected the Ohmic diffusion of the mean field. The helicity
equation for the mean magnetic field can be derived from equation
(\ref{eq:dadt0}):
\begin{equation}
\frac{\partial} {\partial t} \left( {\bf A}_0 \cdot {\bf B}_0 \right)
+ \bnabla \cdot \left[ {\bf A}_0 \times ( {\bf v}_0^\prime \times
{\bf B}_0 + {\cal E}_0 + \bnabla \Phi_0 ) \right] =  2 {\cal E}_0
\cdot {\bf B}_0 . \label{eq:heli1}
\end{equation}
On the other hand, applying the smoothing operator directly to
equation (\ref{eq:heli}) with $\eta \approx 0$, we obtain:
\begin{equation}
\frac{\partial} {\partial t} \left( < {\bf A} \cdot {\bf B} > \right)
+ \bnabla \cdot \left[ {\bf A}_0 \times \left( {\bf v}_0^\prime \times
{\bf B}_0 + {\cal E}_0 + \bnabla \Phi_0 \right) + {\cal H}_0 \right]
= 0 , \label{eq:heli2}
\end{equation}
where ${\cal H}_0$ is the contribution of the fluctuations to the
helicity flux:
\begin{equation}
{\cal H}_0 \equiv \ < {\bf A}_1 \times ( - c {\bf E}_1 + \bnabla
\Phi_1 ) > .
\end{equation}
We assume that the large-scale structures dominate the helicity
budget, so that the helicity of the total field can be approximated by
the helicity of the mean field, $< {\bf A} \cdot {\bf B} > \approx
{\bf A}_0 \cdot {\bf B}_0$ \citep[][]{Boozer1986, Bhattacharjee1986,
Vishniac2001}. Subtracting equation (\ref{eq:heli1}) from equation
(\ref{eq:heli2}) yields the following relationship between
${\cal E}_0$ and ${\cal H}_0$:
\begin{equation}
2 {\cal E}_0 \cdot {\bf B}_0 = - \bnabla \cdot {\cal H}_0 .
\label{eq:2EB}
\end{equation}
The electromotive force can be written as a sum of components parallel
and perpendicular to the mean magnetic field, ${\cal E}_0 =$
${\cal E}_{0,\parallel} + {\cal E}_{0,\perp}$. The parallel component
is given by equation (\ref{eq:2EB}), and the perpendicular component
can be written in terms of a transport velocity ${\bf u}_0$, which is
defined by ${\bf u}_0 \equiv ( {\bf B}_0 \times {\cal E}_{0,\perp})
/ B_0^2$. Then the electromotive force is given by
\begin{equation}
{\cal E}_0 = {\bf u}_0 \times {\bf B}_0 - \frac{ \bnabla \cdot
{\cal H}_0 } {2 B_0^2} {\bf B}_0 . \label{eq:E0}
\end{equation}

We assume that the magnetic field evolves slowly in response to
hyperdiffusion and changes in the boundary conditions. Furthermore,
we assume that the magnetic pressure is much larger than the plasma
pressure, so that the mean magnetic field evolves through a series of
force-free equilibrium states, $( \bnabla \times {\bf B}_0) \times
{\bf B}_0 \approx 0$. The force-free condition can also be written as
\begin{equation}
\bnabla \times {\bf B}_0 = \alpha_0 {\bf B}_0 , \label{eq:fff0}
\end{equation}
where $\alpha_0$ is the torsion parameter of the mean field.
For a linear force free field ($\alpha_0$ = constant) the helicity
flux ${\cal H}_0$ should vanish because there is no magnetic free
energy available to drive the magnetic fluctuations. Therefore,
following \citet{Boozer1986} and \citet{Bhattacharjee1986} we assume
that ${\cal H}_0$ is proportional to the gradient of $\alpha_0$:
\begin{equation}
{\cal H}_0 = - 2 \eta_4 B_0^2 \bnabla \alpha_0 , \label{eq:hyper}
\end{equation}
where $\eta_4$ is the so-called hyperdiffusivity.
Inserting expression (\ref{eq:hyper}) into equations (\ref{eq:E0})
and (\ref{eq:dadt0}) yields the following diffusion equation for
the mean magnetic field:
\begin{equation}
\frac{\partial {\bf A}_0} {\partial t} = {\bf v}_0 \times {\bf B}_0
+ \frac{{\bf B}_0 } {B_0^2} \bnabla \cdot ( \eta_4 B_0^2
\bnabla \alpha_0 ) - \bnabla \Phi_0 , \label{eq:dadt1}
\end{equation}
where ${\bf v}_0 \equiv {\bf v}_0^\prime + {\bf u}_0$ is the total
transport velocity. The second term on the right-hand side of
equation (\ref{eq:dadt1}) describes the effect of the magnetic
fluctuations in diffusing the mean magnetic field. The effect is such
that the magnetic helicity of the system is conserved. Hyperdiffusion
describes the tendency of the magnetic field to relax to a state of
constant $\alpha_0$ \citep[][]{Taylor1974}. The heating of the coronal
plasma resulting from Taylor relaxation was first modeled by
\citet{Heyvaerts1984}.

The energy equation for the mean magnetic field can be derived from
equations (\ref{eq:dadt0}), (\ref{eq:E0}) and (\ref{eq:fff0}):
\begin{equation}
\frac{\partial} {\partial t}
\!
\left( \frac{B_0^2}{8 \pi} \right) +
\bnabla \cdot \left( \frac{c}{4 \pi} {\bf E}_0 \times {\bf B}_0
\right) = - \frac{c}{4 \pi} {\bf E}_0 \cdot ( \bnabla \times {\bf B}_0
) = - \frac{\alpha_0}{8 \pi} \bnabla \cdot {\cal H}_0 ,
\label{eq:ener1}
\end{equation}
and inserting equation (\ref{eq:hyper}) we can write
\begin{equation}
\frac{\partial} {\partial t} \left( \frac{B_0^2 }{8 \pi} \right) +
\bnabla \cdot \left( \frac{B_0^2}{4 \pi} {\bf v}_{0,\perp} - 
\eta_4 \frac{B_0^2} {4 \pi} \alpha_0 \bnabla \alpha_0 \right) =
- \eta_4 \frac{B_0^2} {4 \pi} | \bnabla \alpha_0 |^2 .
\label{eq:ener2}
\end{equation}
where ${\bf v}_{0,\perp} = {\bf v}_{0,\perp}^\prime + {\bf u}_0$ is
the total transport velocity. The quantity on the RHS of equation
(\ref{eq:ener2}) is negative and can be interpreted as the rate at
which mean magnetic energy is converted into heat
\citep[][]{Boozer1986, Bhattacharjee1986}. Assuming there is no inflow
of energy at the boundaries, the total magnetic energy decreases
monotonically with time:
\begin{equation}
\frac{d}{dt} \left( \int \frac{B_0^2}{8 \pi} ~ dV \right) =
- \int \eta_4 \frac{B_0^2}{4 \pi} | \bnabla \alpha_0 |^2 ~ dV < 0 .
\label{eq:Rt}
\end{equation}

\section{Helicity Flux in Stochastic Magnetic Fields}

The above theory predicts a particular form of the mean-field
diffusion equation, but does not explain why the helicity flux
${\cal H}_0$ is proportional to $\bnabla \alpha_0$, nor does it
predict the value of the hyperdiffusivity $\eta_4 ({\bf r},t)$
contained in equation (\ref{eq:hyper}). To estimate this quantity,
a detailed understanding of the processes that drive the magnetic
fluctuations would be required. \citet{Strauss1988} assumed the
fluctuations are driven by tearing mode instability, but in the
present paper we make no assumption about the nature of the
instability. We only assume that (1) the driving process results in
stochastic magnetic field lines \citep[][]{Lazarian1999}, as
illustrated in Fig.~\ref{fig1}; (2) reconnection occurs intermittently
at many sites within this stochastic field; (3) the reconnection
events result in Alfv\'{e}nic perturbations that travel along the
stochastic field lines until they are damped as a result of turbulent
cascade. Therefore, the magnetic field is not completely force free.
In a non-force-free field, the magnetic torsion parameter $\alpha$ can
be defined as $\alpha \equiv ( {\bf B} \cdot \bnabla \times {\bf B} )
/ B^2$, and we assume that this quantity varies in a time-dependent
way along the field lines. The topology of the field is continually
changing as a result of the reconnection processes, with new magnetic
connections being formed and other connections being severed.

Consider a small volume near the edge of the coronal flux rope where
the mean torsion $\alpha_0 ({\bf r})$ can be approximated as a linear
function of position. Let $z$ be the coordinate perpendicular to the
surfaces of constant $\alpha_0$ (we assume $d \alpha_0 /dz > 0$). We
assume that at time $t = 0$ magnetic reconnection occurs at some
localized site, creating a new magnetic connection between two
neigboring surfaces $z$ = constant. Let $P_1$ and $P_2$ be two
points on the newly formed field line, such that $P_1$ is located
on $z = z_1$ and $P_2$ is located on $z = z_2 > z_1$. The
displacement $\Delta s$ between these two points as measured along the
field line is assumed to be much larger than the distance $\Delta z$
between the two surfaces. The displacements are given by
\begin{equation}
\Delta z \sim \lambda_\perp , ~~~~~
\Delta s \sim \lambda_\parallel , \label{eq:disp}
\end{equation}
where $\lambda_\parallel$ and $\lambda_\perp$ are the parallel- and
perpendicular length scales of the turbulent motions (relative to the
mean magnetic field). The torsion parameters $\alpha (P_1)$ and
$\alpha (P_2)$ at the two points differ by an amount
\begin{equation}
\Delta \alpha \equiv \alpha (P_2) - \alpha (P_1) \approx
\alpha_0 (z_2) - \alpha_0 (z_1) \approx \Delta z \frac{d \alpha_0}
{dz} . \label{eq:dalpha}
\end{equation}
We assume that this difference in torsion results in the launch of
Alfv\'{e}n waves that travel in both directions along the newly formed
field line. In Appendix A we show that these waves carry a certain
amount of magnetic helicity. The net helicity flux due to the waves
can be estimated by inserting expression (\ref{eq:dalpha}) into
equation (\ref{eq:flux}):
\begin{equation}
{\cal H}_\parallel \approx - f_0 \lambda_\perp^2 v_A B_0^2 ~ \Delta z
\frac{d \alpha_0} {dz} ,
\label{eq:fluxp}
\end{equation}
where $v_A$ is the Alfv\'{e}n speed, and we assume that the radius $R$
of the newly formed flux tube is determined by the outer scale of the
turbulence, $R \approx \lambda_\perp$. The quantity $f_0$ is a
dimensionless constant, and we estimate it to be rather small,
$f_0 \approx 0.01$ (see Appendix). The $z$-component of the helicity
flux is
\begin{equation}
{\cal H}_{0,z} \approx  {\cal H}_\parallel \frac{\Delta z} {\Delta s}
\approx - f_0 \lambda_\perp^4 \lambda_\parallel^{-1} v_A
B_0^2 \frac{d \alpha_0} {dz} , \label{eq:fluxx}
\end{equation}
where we take into account the angle of the field lines with respect
to the $z$-direction, and we use equations (\ref{eq:fluxp}) and
(\ref{eq:disp}). This shows that the helicity flux ${\cal H}_0$ due to
the fluctuations is proportional to the local gradient of $\alpha_0$,
as assumed in equation (\ref{eq:hyper}). Furthermore, we obtain the
following estimate for the hyperdiffusivity:
\begin{equation}
\eta_4 \approx f_0 \lambda_\perp^4 \lambda_\parallel^{-1} v_A .
\label{eq:eta4}
\end{equation}
Note that $\eta_4$ depends strongly on the perpendicular length scale
of the turbulent motions associated with the reconnection in the
stochastic magnetic field.

\section{Coronal Heating}

The coronal magnetic field is anchored in the photosphere where the
field is highly fragmented \citep[][]{Stenflo1989}. The photospheric
field consists of vertically oriented, kilo-Gauss {\it flux tubes}
separated by nearly field-free plasma. These flux elements are
continually buffeted by convective flows on the scale of the solar
granulation, which results in random twisting of the flux tubes and
a random walk of the flux elements over the solar surface
\citep[e.g.,][]{Berger1996, Nisenson2003}. Both types of motions
cause disturbances that propagate upward along the field lines in the
form of Alfv\'{e}n waves \citep[e.g.,][]{Cranmer2005}. In coronal
loops, such disturbances produce small-scale twisting and braiding of
the coronal field lines and associated field-aligned electric currents
\citep[][] {Parker1972}. \citet{Schrijver2007} argues that the
observed widths of coronal loops are determined by these footpoint
motions. The random walk of the footpoints is predicted to drive a
cascade of magnetic energy from large to small spatial scales
\citep[][]{vB1985, vB1986, vB1988, Craig2005}. The existence of such
a cascade is confirmed by results from 3D numerical simulations of
the Parker problem with incompressible boundary flows \citep[e.g.,][]
{Mikic1989, Hendrix1996, Hendrix+etal1996, Galsgaard1996,
Rappazzo2007}. However, the twisting motions of the flux tubes may be
more important than the lateral motions, as the twist velocity is
amplified by the expansion of the flux tubes \citep[][]{vB1986,
Mandrini2000}. At the smallest scales in the corona, impulsive
reconnection events (``nanoflares'') play an important role in the
release of the magnetic energy \citep{Parker1988, Lu1991}. The
actual dissipation may be a multi-step process involving reconnection
outflows, waves, and energetic particles \citep[][] {Longcope2004}.
\citet{Gudiksen2005} have modeled the coronal heating that ultimately
results from granule-scale footpoint motions.

In this paper we assume that the coronal heating rate $\epsilon$ has
contributions from two different sources: (1) disspation of energy
that was directly injected into the corona by granule-scale
footpoint motions, as described above; (2) dissipation of energy from
a large-scale coronal flux rope via the process of hyperdiffusion.
The former uses energy propagated upward along the field lines, while
the latter draws its energy from large-scale magnetic shear already
present in the corona. The direct contribution to the heating is
approximately given by
\begin{equation}
\epsilon_{\rm direct} (s) \approx \tilde{q} ~ \frac{B_0^2 (s) u^2 (s)
\taup} {4 \pi L^2} , \label{eq:eps_direct1}
\end{equation}
where $s$ is the position along the coronal loop,
$\epsilon_{\rm direct} (s)$ is the heating rate per unit volume,
$B_0 (s)$ is the magnetic field strength, $u(s)$ is the velocity of
the perpendicular motions, $\taup$ is the correlation time of the
photospheric motions, $L$ is the loop length, and $\tilde{q}
\approx 3$. This expression is a generalization of equation (61) in
\citet{vB1986}, using a magnetic Reynolds number $R_m \sim 10^{10}$.
Equation (\ref{eq:eps_direct1}) assumes that the Alfv\'{e}n-travel
time along the coronal loop is short compared to the correlation time,
$L /v_A < \taup$. The perpendicular velocity is given by
\begin{equation}
u(s) \approx \uphot \sqrt{ \frac {\bphot} {B_0 (s)} } ,
\label{eq:us}
\end{equation}
where $\uphot$ is the root-mean-square velocity in the photosphere,
$\bphot$ is the photospheric field strength, and we assume that the
magnetic flux associated with the perturbations is constant along the
field lines from the base of the photosphere into the corona.
Combining the above equations, the direct heating rate is
\begin{equation}
\epsilon_{\rm direct} (s) \approx \tilde{q} ~  \frac{\bphot \uphot^2
\taup} {4 \pi L^2} B_0 (s) , \label{eq:eps_direct}
\end{equation}
similar to equation (67) of \citet{vB1986}. In the present paper we
use $\bphot = 1$ kG, $\uphot = 0.35$ $\rm km ~ s^{-1}$, and $\taup =
600$ s, which yields a photospheric diffusion constant $D = 92$
$\rm km^2 ~ s^{-1}$ [see equation (62) of \citet{vB1986}], consistent
with the observed dispersal coefficient that characterizes the
granular random walk up to several hours \citep{Hagenaar1997}.

The heating rate due to hyperdiffusion follows from equations
(\ref{eq:ener2}) and (\ref{eq:eta4}):
\begin{equation}
\epsilon_{\rm hyper} (s) \equiv \eta_4 \frac{B_0^2}{4 \pi} | \bnabla
\alpha_0 |^2 \approx f_0 \frac{B_0^2 (s) \lambda_\perp^4 (s)}
{4 \pi L} v_A (s) | \bnabla \alpha_0 |^2 , \label{eq:eps_hyper1}
\end{equation}
where we assume that the parallel length scale of the fluctuations
is determined by the loop length, $\lambda_\parallel \sim L$. 
The perpendicular length scale $\lambda_\perp (s)$ of the turbulence
is assumed to vary along the field lines in a manner consistent
with magnetic flux conservation:
\begin{equation}
\lambda_\perp (s) \approx \lenp \sqrt{ \frac {\bphot} {B_0 (s)} } ,
\label{eq:lperp}
\end{equation}
where $\lenp$ is the corresponding length scale in the photosphere.
However, the turbulence is not driven by the footpoint motions,
therefore $\lenp$ does not necessarily reflect any property of the
photospheric velocity field. We cannot provide a more detailed
expression for $\lenp$ because the nature of the physical processes
that produce the stochastic field are still unclear; instead, we
treat $\lenp$ as a free parameter of the model. Combining the above
expressions, we find
\begin{equation}
\epsilon_{\rm hyper} (s) \approx f_0 \frac{\bphot^2 \lenp^4}
{4 \pi L} v_A (s) | \bnabla \alpha_0 |^2 , \label{eq:eps_hyper}
\end{equation}
The total heating rate is assumed to be the sum of expressions
(\ref{eq:eps_direct}) and (\ref{eq:eps_hyper}):
\begin{equation}
\epsilon (s) = \epsilon_{\rm direct} (s) + \epsilon_{\rm hyper} (s) .
\label{eq:eps}
\end{equation}
In the next section we use these expressions to compute the coronal
temperature and density, which provides a link to solar observations.

\section{Active Region Model}

\citet{Bobra2008} developed empirical models for NLFFFs in two active
regions observed with TRACE. Our initial goal was to compute
temperatures and densities for these models. However, we found that
small deviations from the force-free condition can cause significant
errors in the distribution of $\bnabla \alpha_0$ along the field
lines, so that the heating rate $\epsilon_{\rm hyper} (s)$ cannot yet
be accurately computed from such empirical models. Therefore, in
the present paper we apply the above expressions for the heating rate
to a semi-analytic model of an active region. The model describes the
{\it mean} magnetic field ${\bf B}_0 ({\bf r})$ in the corona;
fluctuating fields are not modeled, nor do we consider the evolution
of the mean field.

\begin{figure}
\epsscale{1.15}
\plotone{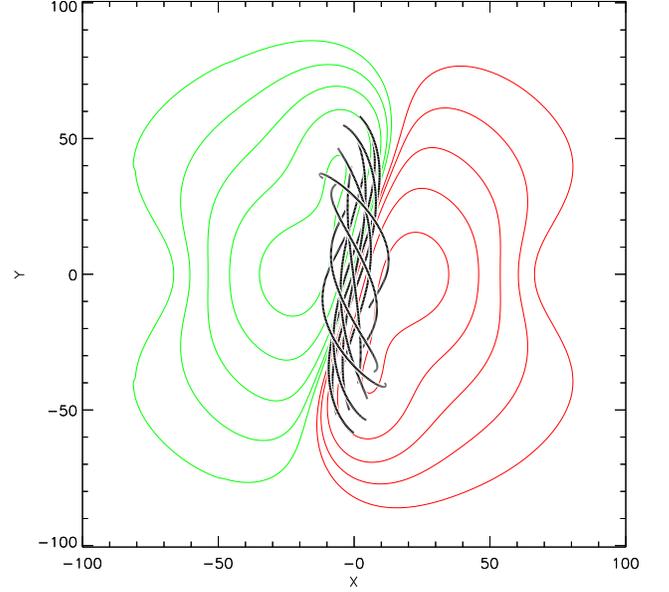}
\caption{Non-linear force-free field model of an active region
containing a magnetic flux rope. The contours show the flux
distribution on the photosphere ({\it red}: positive flux; {\it
green}: negative flux). The black lines show selected field lines in
the flux rope.}
\label{fig2}
\end{figure}

The NLFFF model is constructed as follows. We use a Cartesian
coordinate system $(x,y,z)$ with $z$ being the height above the
coronal base. Following \citet{Titov1999}, we assume that the magnetic
field has cylindrical symmetry with respect to an axis that lies
horizontally below the coronal base at $y=0$, $z=-R$. Let $(r,\phi,x)$
be a cylindrical coordinate system with $r$ the distance from the
cylinder axis and $x$ the distance along the axis. We consider a
finite domain $r = [R,R_m]$ and $x = [-x_m,x_m]$, where the inner
radius of the domain is given by $R$, and $x_m \gg R$. We initially
use dimensionless units such that $R=1$. The force free field is
written as:
\begin{equation}
B_{0,r} = - \frac{1}{r} \frac{\partial A} {\partial x} , ~~~~
B_{0,x} = \frac{1}{r} \frac{\partial A} {\partial r} , ~~~~
B_{0,\phi} = \frac{B(A)}{r} ,
\end{equation}
where $A(r,x)$ is the flux function, and $B(A)$ describes the
azimuthal component of the field. The flux function satisfies the
following non-linear equation:
\begin{equation}
r \frac{\partial} {\partial r} \left( \frac{1}{r} \frac{\partial A}
{\partial r} \right) + \frac{\partial^2 A} {\partial x^2} +
\alpha_0 (A) B(A) = 0 , \label{eq:cfff}
\end{equation}
where $\alpha_0 (A) \equiv dB/dA$. In this paper we use boundary
conditions such that $A(1,x)=$ $0.8 \exp (- x^2)$ at the inner
boundary, and $A=0$ at the outer boundaries. The azimuthal field is
given by
\begin{equation}
B(A) = \beta [ 1 - \exp ( - A^5 ) ] .
\end{equation}
where $\beta$ is a parameter describing the magnitude of the azimuthal
field. Equation (\ref{eq:cfff}) is solved using a relaxation technique.
In essense, we solve the following time-dependent equation:
\begin{equation}
\frac{\partial A} {\partial t} = D \left[
r \frac{\partial} {\partial r} \left( \frac{1}{r} \frac{\partial A}
{\partial r} \right) + \frac{\partial^2 A} {\partial x^2} +
\alpha_0 (A) B(A) \right] , \label{eq:cfff2}
\end{equation}
until a stationary solution $A(r,x)$ is obtained (here $D$ is a
dimensionless diffusion constant). In each time step, $\beta$ is
adjusted such that the total azimuthal current has a fixed value,
$\int \int \alpha_0 B ~ dx dz = 2.5$. Once a stationary solution is
obtained, the spatial scale is changed to $R = 30$ Mm, and the
magnetic field is rescaled by a factor 100, producing a maximum value
of $|B_{0,\phi}|$ of 55 G. The quantities $B_{0,r}$, $B_{0,x}$ and
$B_{0,\phi}$ are computed on a grid of $200 \times 200$ points in the
$(r,x)$ plane. The magnetic vector ${\bf B}_0 (x,y,z)$ in the
cartesian frame can then be found by evaluating these quantities at
$r = \sqrt{y^2+(R+z)^2}$. Figure \ref{fig2} shows the magnetic flux
distribution at the coronal base ($z=0$) and selected field lines in
the flux rope. The stability of the \citet{Titov1999} model has been
considered by \citet{Torok2004}. They find that the $m=1$ kink mode is
stable provided the average twist in the flux rope is less than $3.5
\pi$.  This condition is satisfied in the present case, so we assume
that the field shown in Fig.~\ref{fig2} is stable.

The NLFFF model provides the spatial distribution of $\alpha_0
({\bf r})$, which is needed to compute the heating rate $\epsilon$.
We trace a large number of field lines that cross the vertical plane
$y = 0$, and solve the energy balance equation for each field line. We
use a modified version of the iterative method described by
\citet{Schrijver+vB2005}. The code assumes steady-state heating, and
includes thermal conduction and optically thin radiative losses.  The
contribution to the heating from hyperdiffusion is determined from
equation (\ref{eq:eps_hyper}) with the Alfv\'{e}n speed $v_A (s)$
taken from the previous iteration. We assumed that the perpendicular
length scale of the turbulence is characterized by $\lenp = 10^3$ km.
The chromosphere-corona transition region (TR) is assumed to be
located at $z = 0$, where the temperature $T = 2 \times 10^4$ K.
The solution of the energy balance equation yields the temperature
$T(s)$ and electron density $n_e (s)$ as functions of position $s$
along a field line.

\begin{figure}
\epsscale{1.19}
\plotone{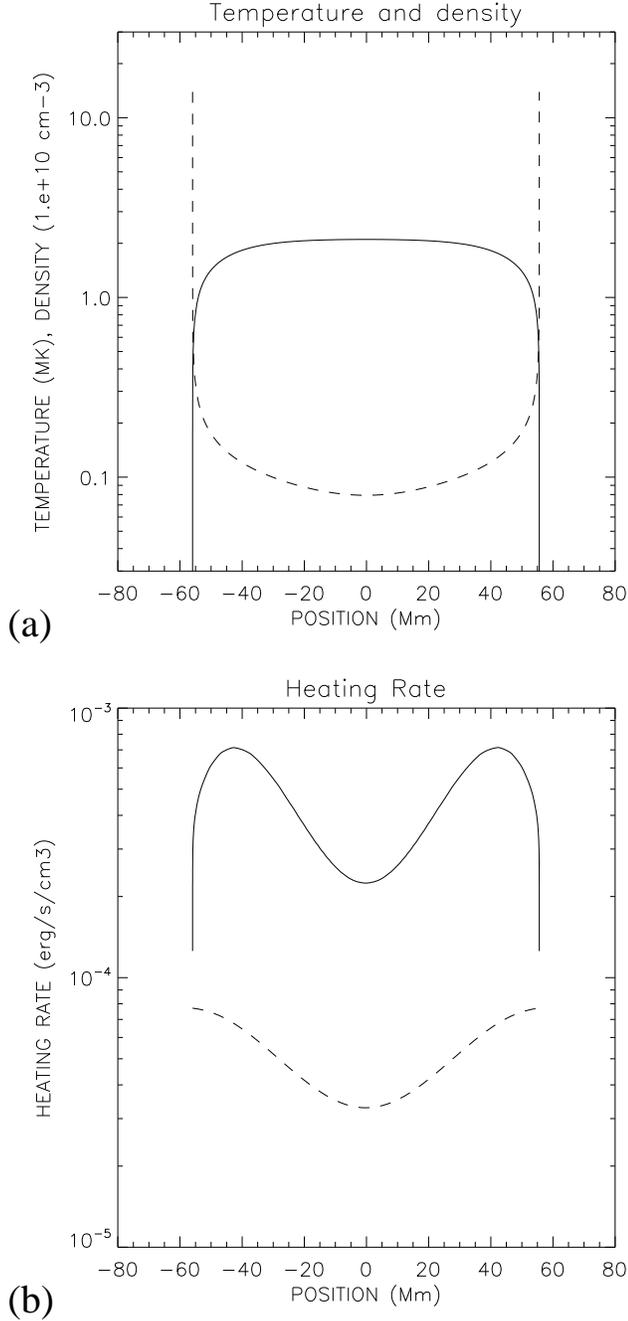}
\caption{Results from modeling the energy balance of the plasma
along a particular magnetic field line located at the edge of the
coronal flux rope where $| \bnabla \alpha_0 |$ is largest.
(a) Temperature $T(s)$ ({\it full} curve) and electron density
$n_e (s)$ ({\it dashed} curve) as functions of position $s$ along
the field line. (b) Total heating rate $\epsilon (s)$ ({\it full}
curve), which is a sum of hyperdiffusive heating and a ``direct''
contribution from photospheric footpoint motions. The latter is
shown by the {\it dashed} curve.}
\label{fig3}
\end{figure}

Figure \ref{fig3} shows modeling results for one field line at the
edge of the coronal flux rope. Starting at $x=y=0$ and $z=43$ Mm, the
field line is traced forward and backward through the 3D magnetic
model until the level $z=0$ is reached. The temperature $T(s)$ and
density $n_e (s)$ along this field line are shown in Fig.~\ref{fig3}a,
and the heating rate $\epsilon (s)$ is shown in Fig.~\ref{fig3}b
({\it full} curve). The dashed curve in Fig.~\ref{fig3}b shows the
``direct'' contribution to the heating from photospheric footpoint
motions. Note that this contribution is relatively small, so for this
particular field line the heating is dominated by hyperdiffusion.
However, for other field lines in the NLFFF model (e.g., along the
axis of the flux rope) the hyperdiffusive heating rate is smaller than
the direct heating rate.

\begin{figure}
\epsscale{1.15}
\plotone{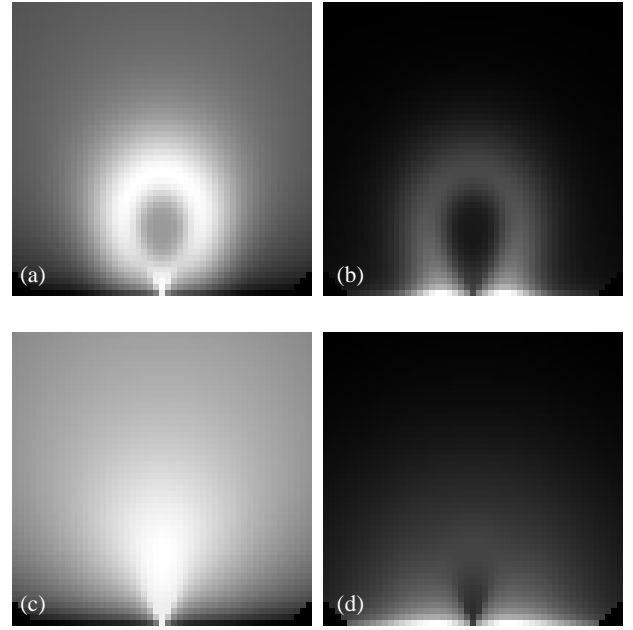}
\caption{Coronal models for an active region containing a magnetic
flux rope. The upper panels show results for a model with both direct
and hyperdiffusive heating: (a) temperature $T(x,z)$ in a vertical
plane $y = 0$ intersecting the flux rope; (b) electron density $n_e
(x,z)$ in the same vertical plane. The height and width of these
greyscale images correspond to 100 Mm on the Sun. The maximum
temperature in the model is 2.11 MK, and the maximum density is $2.7
\times 10^9$ $\rm cm^{-3}$. The bottom panels show results for a model
without hyperdiffusion: (c) temperature $T(x,z)$, and (d) electron
density $n_e (x,z)$. This model has lower peak temperature and density
(maximum values are 1.42 MK and $1.0 \times 10^9$ $\rm cm^{-3}$,
respectively). In both models the minimum temperature is 0.37 MK, and
the minimum density is $1.8 \times 10^7$ $\rm cm^{-3}$. Maximum values
are shown as white; minimum values are black.}
\label{fig4}
\end{figure}

We computed temperature and density profiles for a large number of
field lines that cross the vertical plane $y = 0$. Figure \ref{fig4}
shows the temperature $T(x,z)$ and density $n_e (x,z)$ in the plane
$y = 0$ for two different heating models. The top panels show the
results when both direct and hyperdiffusive heating are applied
together ($\lenp = 10^3$ km). Note that the temperature and density
are enhanced in a thick shell where the gradient of $\alpha_0$ is
largest. The density distribution shows a ``cavity'' that is round
at the top and V-shaped at the bottom. The temperature is also
enhanced at the bottom of the V-shaped cavity because the field lines
in this region are relatively long, so the conductive cooling is
reduced.

The bottom panels of Figure \ref{fig4} show the temperature and
density distributions in the $y=0$ plane when only the direct heating
is applied (i.e., no hyperdiffusion). The highest temperatures now
occur within the flux rope where the field lines are longest and
conductive losses are weakest. The lowest densities occur in the
V-shaped lower part of the flux rope, but the density in the upper
part is nearly the same as in the surroundings, and there is no
dome-shaped cavity.

On the Sun, cavities are frequently observed in the K-corona emission,
and these cavities have been linked to the presence of coronal flux
ropes \citep[see][and references therein] {Gibson+etal2006b}.
We suggest that the observed cavities are evidence for a spatial
distribution of coronal heating that is consistent with hyperdiffusion.

\section{Discussion and Conclusions}

We propose a new mechanism for coronal heating relevant for magnetic
structures in which the large-scale magnetic field is highly sheared
and/or twisted. We assume that the magnetic field in such coronal flux
ropes is stochastic in nature \citep[][]{Lazarian1999}, and contains
small-scale perturbations superposed on a mean magnetic field. The
magnetic free energy associated with the flux rope can be converted
into heat via small-scale reconnection processes.  We describe this
process in terms of {\it hyperdiffusion}, a type of magnetic diffusion
in which the magnetic helicity is conserved.  The effect of
hyperdiffusion on the mean field depends on the gradient of the mean
torsion parameter $\alpha_0 ({\bf r},t)$.  Therefore, hyperdiffusion
describes the tendency of a magnetic system to relax to a Taylor state
\citep[][]{Taylor1974}.

In section 3 we consider helicity transport in a stochastic magnetic
field, and we show that small-scale reconnection events produce a
net helicity flux ${\cal H}_0$ consistent with hyperdiffusion. We
derive an expression for the hyperdiffusivity $\eta_4$ in terms of
the parallel and perpendicular length scales of the magnetic
perturbations. The calculation is based on a simple model for the
Alfv\'{e}n waves produced by reconnection events (Appendix A). We
find that the hyperdiffusivity depends strongly on the perpendicular
length scale of the turbulence, $\eta_4 \propto \lenp^4$.
The spatially averaged heating rate $\epsilon_{\rm hyper}$ due to
hyperdiffusion is proportional to $\eta_4 | \bnabla \alpha_0 |^2$.
This heating is distinct from the ``direct'' heating due to
small-scale random footpoint motions. In section 4 we derive an
expression for the total heating rate $\epsilon$, which is a sum of
direct and hyperdiffusive contributions.

Our analysis of the dissipation process is very approximate and makes
many simplifying assumptions. For example, we cannot identify a
specific mechanism by which the stochastic magnetic field is produced
and maintained. We simply treat the perpendicular length scale $\lenp$
of the fluctuating field as a free parameter to be determined from
observations or more complete theories of turbulent reconnection.
Also, it is unclear exactly what drives the small-scale reconnection
events necessary for hyperdiffusion. The photospheric magnetic field
is highly fragmented, and there are separatrix surfaces in the corona
corresponding to the interfaces between photospheric flux tubes.  We
speculate that the reconnection events may occur at these separatrix
surfaces.

In section 5 we develop a 3D model of an active region containing a
magnetic flux rope, and we compute the temperature and density
distributions with and without hyperdiffusion. The results show that
hyperdiffusion can have a significant effect on the temperature
structure, provided the perpendicular length scale of the magnetic
fluctuations $\lenp \sim 10^3$ km at the coronal base. If $\lenp$
were of order 100 km or less, the hyperdiffusive heating would not
have a significant impact on the total heating rate for any field
line. For the NLFFF model considered here, the heating is concentrated
in a thick shell surrounding the flux rope. The heating at the axis
of the flux rope is weaker, producing a ``cavity'' similar to the
coronal cavities observed on the Sun. 

At present there is no direct observational evidence to confirm or
refute the existence of stochastic magnetic fields with perpendicular
scales $\sim 10^3$ km. Therefore, it is unclear exactly how coronal
flux ropes are heated. Coronal images obtained with the {\it Solar
Dynamics Observatory} (SDO) may provide new observational constraints
on stochastic fields and the physical processes by which coronal flux
ropes are heated.

\acknowledgements
We thank Amitava Bhattacharjee for discussions about hyperdiffusion.
This work was inspired by observations from Hinode/XRT. The work
was supported by NASA LTSA grant NNG04GE77G, and by NASA LWS grant
NNX06AG95G.

\appendix

\section{Transport of Helicity by Alfv\'{e}n Waves in a Stochastic Field}

The transport of magnetic helicity plays an important role in the
theory of hyperdiffusion. Here we consider a simple model for helicity
transport due to reconnection events within the stochastic field.
The average time $T$ between reconnection events on a given flux tube
is assumed to be a few times the Alfv\'{e}n travel time along that tube. 
Let us assume that at time $t = 0$ a new section of magnetic field has
just been created by reconnection. The new section connects two points
$P_1$ and $P_2$ that may be located at opposite ends of a coronal
loop. We neglect the loop curvature and approximate the new section as
a cylindrical flux tube with radius $R$. We assume that in the initial
state the plasma in the tube is stationary, but the torsion parameter
$\alpha$ varies along the tube, so the field is not force free.
The variable torsion causes the launch of counter-propagating
Alfv\'{e}n waves that travel in opposite directions along the tube.
We assume that the wave length $\lambda_\parallel$ of the waves is
large compared to the tube radius. The medium surrounding the flux
tube is assumed to be unaffected by the waves.

We use a cylindrical coordinate system $(r,\phi,s)$ where $s$ is the
distance along the tube ($0 \le s \le \Delta s$). The plasma is
assumed to be fixed at the end points $P_1$ ($s=0$) and $P_2$
($s=\Delta s$), so that the waves reflect at these points. The
magnetic field ${\bf B}$ inside the tube is written as
\begin{equation}
{\bf B} (r,s,t) = B_0 ~ \hat{\bf s} + \frac{r}{R} \left[ B_{\phi,0}
+ \delta B_\phi (s,t) \right] \bphi ,
\end{equation}
where $B_0$ is the axial field (constant in space and time),
$B_{\phi,0}$ is the unperturbed azimuthal field at the cylinder wall
($r=R$), and $\delta B_\phi (s,t)$ is the perturbation of the
azimuthal field due to the waves. The torsion parameter $\alpha$ is
defined by $\alpha \equiv ( {\bf B} \cdot \bnabla \times {\bf B} )
/ B^2$, and the perturbation of this quantity is
\begin{equation}
\delta \alpha (s,t) = \frac{2}{R B_0} \delta B_\phi (s,t) .
\end{equation}
The vector potential ${\bf A} (r,s,t)$ is obtained by solving the
equations $\bnabla \times {\bf A} = {\bf B}$ and $\bnabla \cdot {\bf
A} = 0$. Continuity with the surrounding medium requires that
${\bf A} (R,s,t)$ is independent of $s$ and $t$. In the limit
$\lambda_\parallel \gg R$, the solution is given approximately by
\begin{equation}
{\bf A} (r,s,t) \approx \frac{r B_0}{2} \bphi + \frac{R^2-r^2}{2R}
\left[ B_{\phi,0} + \delta B_\phi (s,t) \right] \hat{\bf s} ,
\end{equation}
where $r \le R$ and we neglect a small correction $A_r$ needed to make
$\bnabla \cdot {\bf A}$ exactly zero. The velocity of the waves is
assumed to be purely azimuthal, ${\bf v}(r,s,t) = (r/R) \delta
v_\phi(s,t) \bphi$, where $\delta v_\phi(s,t) $ is the velocity at
the cylinder wall. Inserting this expression into equation
(\ref{eq:Phi2}), we obtain the electric potential:
\begin{equation}
\Phi (r,s,t) \approx - \frac{R^2-r^2}{2R} B_0 \delta v_\phi(s,t) ,
\end{equation}
where we assumed $\eta = 0$ and required $\Phi (R,s,t) = 0$ to ensure
continuity with the surrounding medium. The axial component of the
helicity flux follows from equation (\ref{eq:H}):
\begin{equation}
{\cal H}_\parallel (r,s,t) = \left[ {\bf A} \times ( {\bf v} \times
{\bf B} + \bnabla \Phi ) \right] \cdot \hat{\bf s} =
- \frac{r^2 B_0^2} {R} \delta v_\phi (s,t) .
\end{equation}
The average of ${\cal H}_\parallel$ over the cross-section of the
tube is
\begin{equation}
{\cal H}_\parallel (s,t) = - \frac{R B_0^2} {2} \delta v_\phi (s,t) .
\label{eq:Hpara}
\end{equation}
Note that the helicity flux is independent of the unperturbed
azimuthal field $B_{\phi,0}$.

We now compute the helicity flux for damped Alfv\'{e}n waves.
In reality the damping would likely involve a turbulent cascade of
wave energy to smaller spatial scales, but here we use a simple model
based on an effective ``frictional'' damping. The wave equations are
\begin{eqnarray}
\frac{ \partial ( \delta v_\phi )} {\partial t} & = & \frac{B_0}
{4 \pi \rho_0} \frac{ \partial ( \delta B_\phi )} {\partial s}
- \frac{ \delta v_\phi (s,t)} {\tau} , \\
\frac{ \partial ( \delta B_\phi )} {\partial t} & = & B_0
\frac{ \partial ( \delta v_\phi )} {\partial s} ,
\end{eqnarray}
where $\rho_0$ is a constant density and $\tau$ is the frictional
damping time. The boundary conditions are $\delta v_\phi (0,t) =
\delta v_\phi (\Delta s,t) = 0$, which implies $\partial ( \delta
B_\phi ) / \partial s = 0$ at the boundaries. We assume the plasma
is initially stationary, $\delta v_\phi(s,0) = 0$. The initial
perturbation of the torsion parameter, $\delta \alpha (s,0)$, is
assumed to be a nearly linear function of position:
\begin{equation}
\delta \alpha (s,0) = - \Delta \alpha ~ \sum_{k=0}^K
\frac{4}{(2k+1)^2 \pi^2} \cos \left[ (2k+1) \pi \frac{s}{\Delta s}
\right] \approx \Delta \alpha \left( \frac{s}{\Delta s}
- \frac{1}{2} \right) , \label{eq:init}
\end{equation}
where $\Delta \alpha$ is the perturbation amplitude (we use $K=15$).
The solution of the above equations is
\begin{eqnarray}
\delta v_\phi (s,t) & = & \frac{\Delta B_\phi} {\sqrt{4 \pi \rho_0}}
~ e^{- \gamma t} ~ \sum_{k=0}^K \frac{4}{n^2 \pi^2} \sin \left( n \pi
\frac{s}{\Delta s} \right) ~ \frac{\Omega_n} {\omega_n} ~ \sin (
\omega_n t) , \label{eq:wv} \\
\delta B_\phi (s,t) & = & - \Delta B_\phi ~ e^{- \gamma t}
\sum_{k=0}^K \frac{4}{n^2 \pi^2} \cos \left( n \pi \frac{s}{\Delta s}
\right) \left[ \cos (\omega_n t) + \frac{\gamma}{\omega_n}
\sin (\omega_n t) \right] , \label{eq:wb}
\end{eqnarray}
where $n \equiv 2k+1$; $\Delta B_\phi = B_0 R \Delta \alpha /2$
is the amplitude of the magnetic perturbation; $\gamma = 1/(2 \tau)$
is the effective damping rate; $\omega_n = \sqrt{ \Omega_n^2 -
\gamma^2}$ is the mode frequency; and $\Omega_n \equiv n \pi v_A /
\Delta s$ is the frequency for an undamped mode. The above solution
is valid for $\gamma \le \pi v_A / \Delta s$. The helicity flux at
the mid-point of the tube ($s = \Delta s/2$) follows from equations
(\ref{eq:Hpara}) and (\ref{eq:wv}):
\begin{equation}
{\cal H}_\parallel (\Delta s/2,t) = - f(t) R^2 B_0^2
v_A \Delta \alpha ,
\end{equation}
where
\begin{equation}
f(t) \equiv e^{- \gamma t} ~ \sum_{k=0}^K \frac{(-1)^k} {(2k+1)^2
\pi^2} ~ \frac{\Omega_{2k+1}} {\omega_{2k+1}} ~
\sin ( \omega_{2k+1} t) .
\end{equation}
In the absence of damping ($\gamma = 0$), the helicity would oscillate
back and forth between the two halves of the tube, and the time
average of ${\cal H}_\parallel$ would vanish. With damping there is a
net transport of helicity from the region with high $\alpha$ to the
region with low $\alpha$:
\begin{equation}
{\cal H}_\parallel \approx - f_0 R^2 B_0^2 v_A \Delta \alpha ,
\label{eq:flux}
\end{equation}
where $f_0$ is the time average of $f(t)$. We assume that the next
reconnection event occurs at a time $T$ which is only a few times the
Alfv\'{e}n travel time, say, $T = 3 \Delta s/v_A$. Then $f_0$ is given
by a time average over the period $0 < t < T$. For a damping time
$\tau$ such that $0.6 < \gamma \Delta s/v_A < \pi$, the time average
of $f(t)$ is given by $f_0 \approx 0.01$. In section 3 we use the
above expressions to estimate the hyperdiffusivity in a stochastic
field.

\end{document}